\documentclass[final]{aipproc}
\layoutstyle{6x9}

\usepackage{amssymb,amsmath}

\newcommand{\myunit}[1]{\; \text{#1} }
\newcommand{\myvec}[1]{\mathbf{#1}}

\newcommand{\reffig}[1]{Fig.~\ref{#1}}
\newcommand{\refeq}[1]{Eq.~(\ref{#1})}
\newcommand{\refcite}[1]{Ref.~\cite{#1}}
\newcommand{\refscite}[1]{Refs.~\cite{#1}}
\newcommand{\refetal}[1]{\emph{et~al.}~\cite{#1}}

\newcommand{\doubleplotscale}{.55}

\begin{document}

\title{Neutrino Interactions with Nuclei}


\author{T.~Leitner}{address={Institut f\"ur Theoretische Physik, Universit\"at Giessen, Germany}}
\author{O.~Buss}{address={Institut f\"ur Theoretische Physik, Universit\"at Giessen, Germany}}
\author{U.~Mosel}{address={Institut f\"ur Theoretische Physik, Universit\"at Giessen, Germany}}
\author{L.~Alvarez-Ruso}{address={Departamento de F\'{\i}sica Te\'orica and IFIC, Universidad de Valencia - CSIC, Spain}}

\begin{abstract}
We investigate neutrino-nucleus collisions at intermediate
energies incorporating quasi-elastic scattering and the excitation
of 13 resonances as elementary processes, taking into account
medium effects such as Fermi motion, Pauli blocking, mean-field
potentials and in-medium spectral functions. A coupled-channel
treatment of final state interactions is achieved with the GiBUU
transport model. Results for inclusive reactions, neutrino- and
electron-induced, as well as for pion production and nucleon
knockout are presented.
\end{abstract}

\maketitle

%%%%%%%%%%%%%%%%%%%%%%%%%%%%%%%%%%%%%%%%%%%%
%% MAINMATTER
%%%%%%%%%%%%%%%%%%%%%%%%%%%%%%%%%%%%%%%%%%%%

%\section{Introduction}

The study of neutrino interactions with nuclei is crucial for
current and future oscillation experiments. The main goal is to
improve our knowledge of the energy fluxes, backgrounds and
detector responses in order to minimize systematic uncertainties.
Most of the experiments are performed on nuclear targets, thus, an
understanding of nuclear effects is essential for the
interpretation of the data.

%\section{Inclusive neutrino-induced reactions}

Here we report on an application to such processes with a model
that has been well tested on other nuclear reactions, such as
heavy-ion collisions, proton- and pion-induced reactions on nuclei
and photonuclear reactions, using the same theoretical input and
the same code: the Giessen BUU model (GiBUU). The model treats the
nucleus as a local Fermi gas of nucleons with the total reaction
rate given by an incoherent sum over all nucleons embedded in a
nuclear medium (impulse approximation). For more details, we refer
the reader to our earlier work~\cite{Leitner:2006,Buss:2007ar}.

At neutrino energies ranging from $0.5-2$ GeV, the relevant
contribution to the cross section is quasi-elastic (QE) scattering
($\nu  N \to l  N'$) and pion production ($\nu  N \to l  \pi N'$).
The latter is dominated by the excitation of the $\Delta$
resonance and its subsequent decay ($l  N \to l' \Delta \to l' \pi
N'$) - however, with increasing neutrino energy also higher
resonances contribute significantly to pion production. The cross
section for QE scattering and resonance excitation on a bound
nucleon is given by
\begin{equation}
\frac{d\sigma_{QE,RES}}{d \omega \; d \Omega} =
\frac{1}{32 \pi^2} \; \frac{|\myvec{k'}|}{ k \cdot p} \;
\mathcal{A}(E',\myvec{p'}) \; C_{CC,NC} \;
L_{\mu \nu} \; H^{\mu \nu}_{QE,RES}\, , \label{eq:cross}
\end{equation}
where we use the following notation: a lepton with four-momentum
$k=(E_\nu,\myvec{k})$ scatters off a nucleon with momentum
$p=(E,\myvec{p})$, going into a lepton with momentum
$k'=(E_{l'},\myvec{k'})$ and a nucleon/resonance with
$p'=(E',\myvec{p'})$. We further define the transferred energy
$\omega=E_\nu-E_{l'}$ and the solid angle $\Omega = \angle
(\myvec{k},\myvec{k'})$. The dynamics of the interaction is
incorporated in the leptonic ($L_{\mu \nu}$) and hadronic ($H^{\mu
\nu}_{QE,RES}$) tensor with the appropriate coupling $C_{CC,NC}$.

For QE scattering, we use the standard expression for $H^{\mu
\nu}_{QE}$ (cf.~e.g.~our earlier work in \refcite{Leitner:2006})
with the BBBA-2005 vector form factors~\cite{Bradford:2006yz} and
a dipole ansatz with $M_A=1$~GeV for the axial ones. For the
resonance excitation, we have considered, besides the dominant
$\Delta$ (P$_{33}$(1232)) resonance, 12 higher resonances, namely
P$_{11}$(1440), D$_{13}$(1520), S$_{11}$(1535), S$_{31}$(1620),
S$_{11}$(1650), D$_{15}$(1675), F$_{15}$(1680), D$_{33}$(1700),
P$_{13}$(1720), F$_{35}$(1905), P$_{31}$(1910) and F$_{37}$(1950).
Following \refscite{Alvarez-Ruso:1997jr,Lalakulich:2006sw}, we
relate the vector form factors to helicity amplitudes for which we
use the results of the recent MAID analysis~\cite{Tiator:2003uu}
while the axial form factors follow a modified dipole ansatz with
an axial coupling obtained with the assumption of PCAC and
pion-pole dominance.

The nucleons are embedded in a nucleus which is treated within a
local Thomas-Fermi approximation as a Fermi gas of nucleons bound
by a density and momentum dependent mean-field. The density
profiles are based on data from electron scattering and
Hartree-Fock calculations. The presence of a momentum-dependent
mean field leads to the appearance of effective masses in
\refeq{eq:cross}. $M$ denotes the effective mass of the incoming
nucleon $N$, defined as $M=M_N+U_{S_N}(\myvec{p},\myvec{r})$,
where $M_N$ denotes its vacuum mass and
$U_{S_N}(\myvec{p},\myvec{r})$ the scalar potential. The spectral
function $\mathcal{A}(E',\myvec{p'})$ includes the effect of the
momentum-dependent potential on the outgoing particle and also
accounts for the in-medium collisional broadening of the outgoing
final states. More details can be found in~\refcite{Buss:2007ar}.

\begin{figure}
        \centerline{
                \includegraphics[scale=\doubleplotscale]{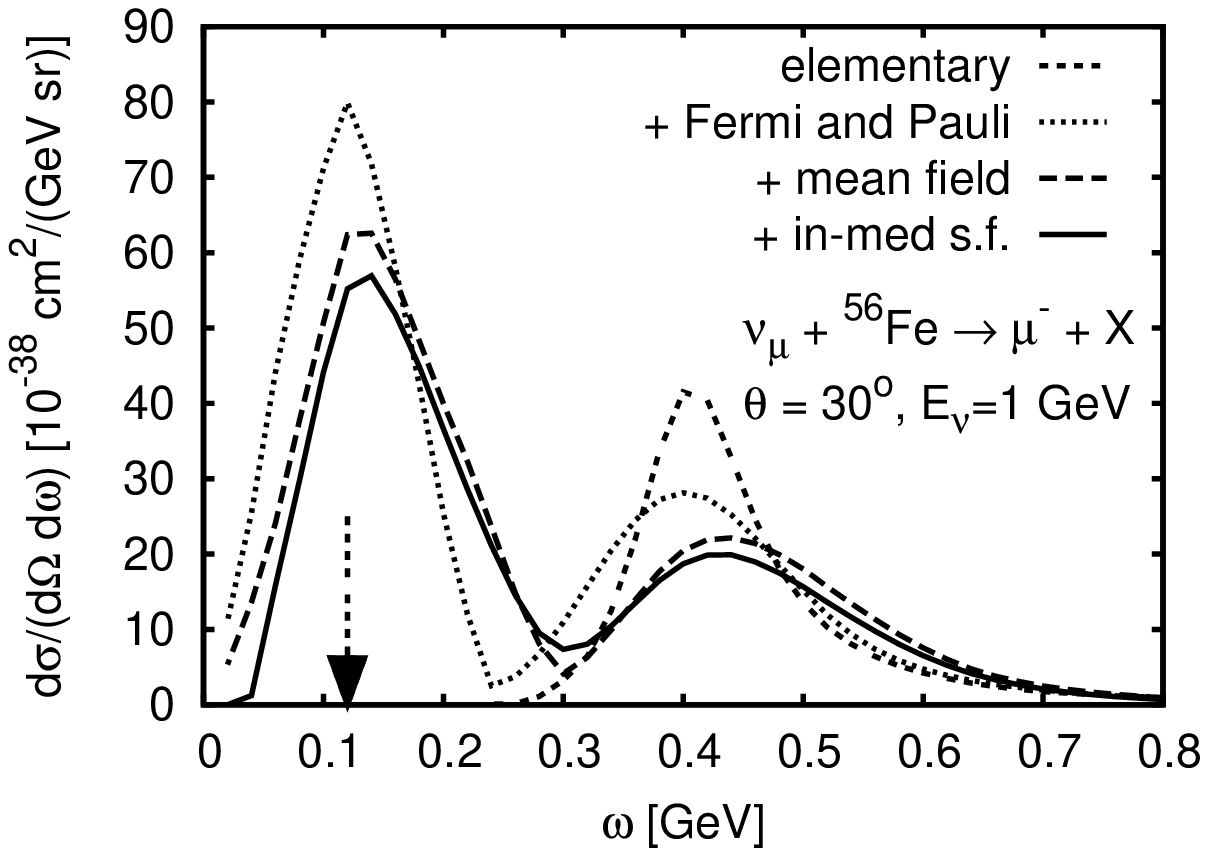}
                \hspace{1em}
                \includegraphics[scale=\doubleplotscale]{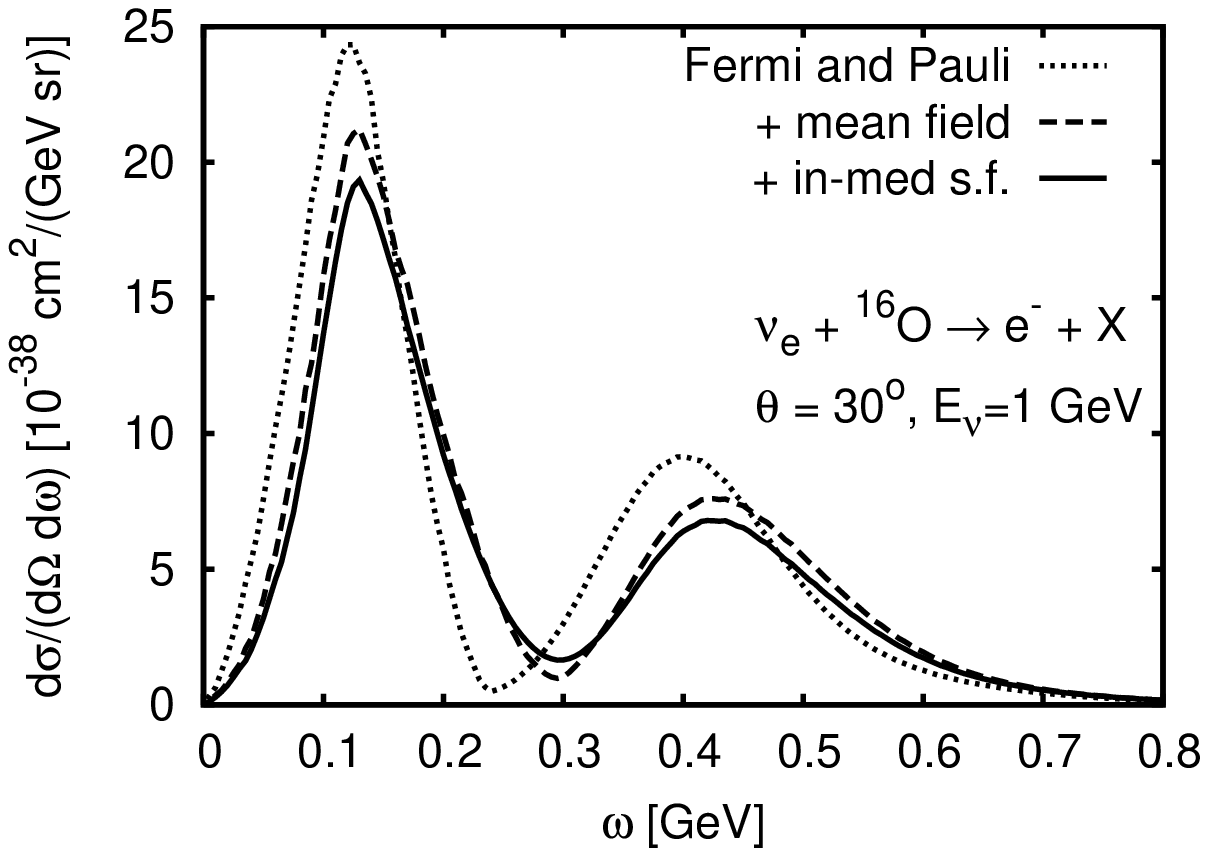}
        } \caption{Inclusive CC cross section $d\sigma / d\omega
d\Omega$ on $^{56}$Fe and $^{16}$O as a function of the energy
transfer $\omega=E_\nu-E_{l'}$ at a $E_\nu=1$ GeV and scattering
angle of $\theta=30^\circ$. The short-dashed line denotes our
result for the free case (left panel only), the dotted line
includes Fermi motion and Pauli blocking only. The long-dashed
line denotes the result, where we take into account also the
binding in a density and momentum dependent mean-field potential.
The solid line includes in addition the in-medium spectral
function (SF). \label{fig:neutrino_doublediff}}
\end{figure}
\begin{figure}
        \centerline{
                \includegraphics[scale=0.85]{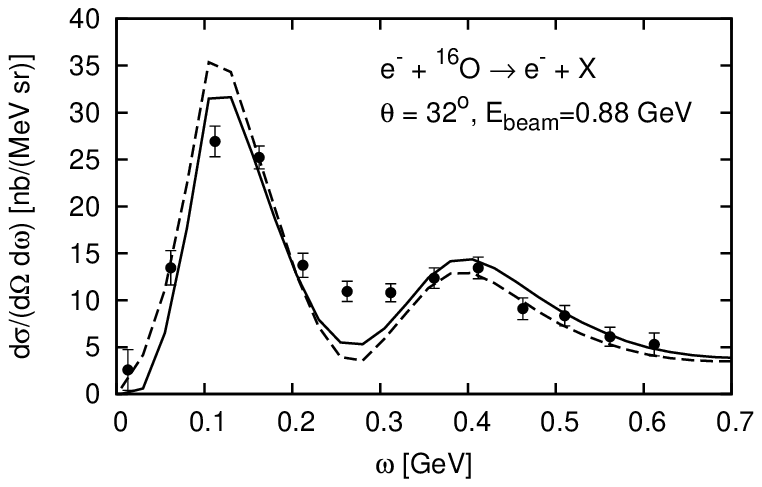}
                \hspace{1em}
                \includegraphics[scale=0.85]{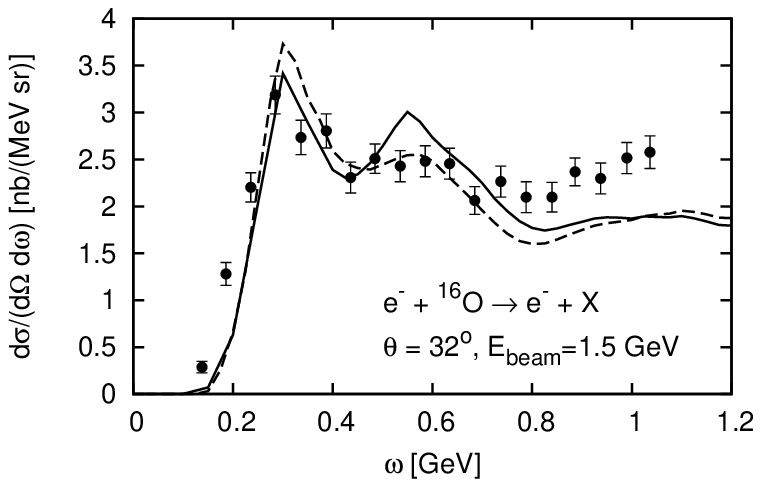}
        } \caption{Same as \reffig{fig:neutrino_doublediff} but
for the electron-induced inclusive reaction
$^{16}$O$\left(e^-,e^-\right)X$ at a fixed electron energy and
scattering angle of $\theta=32^\circ$. The data are taken from
\refcite{Anghinolfi:1996vm}. \label{fig:electro_doublediff}}
\end{figure}
In \reffig{fig:neutrino_doublediff}, we present results for the CC
reaction $^{56}$Fe$\left(\nu_\mu,\mu^-\right)X$ for a neutrino
beam energy of 1 GeV and a lepton scattering angle of $30^\circ$
(left panel) and for $^{16}$O$\left(\nu_e,e^-\right)X$ (right
panel). One observes a broadening and a shift of the QE and pion
peak caused by Fermi motion, the momentum-dependent potential and
the in-medium width of the nucleon and the resonances. As a
necessary check of our calculations we have obtained within the
same model also electron-induced inclusive cross sections. In
these calculations we have omitted the axial parts of the hadronic
currents and included in addition non-resonant single-pion
background~\cite{Buss:2007ar}. The results for
$^{16}$O$\left(e^-,e^-\right)X$ are shown in
\reffig{fig:electro_doublediff} for two different energies. The
overall agreement with the data is very good and comparable to
that in \refcite{Sakuda}. The agreement is improved, in addition
to using a local Fermi gas, by including a mean field and
in-medium spectral functions.

%\section{Exclusive neutrino-induced reactions}

Besides inclusive reactions, also semi-inclusive processes where,
in addition to the outgoing lepton in the $\nu A$ reaction, one or
more pions, nucleons etc.~are detected are experimentally
accessible. In particular for NC reactions one has to rely on the
modeling of these reactions since the outgoing neutrino is not
detected. In our description, we treat the exclusive reaction as a
two step process: once the initial interaction has taken place,
the final state particles are transported out of the nucleus.
These final state interactions (FSI) are implemented by means of
the coupled-channel GiBUU transport model~\cite{Buss:2006yk} based
on the BUU equation
\begin{equation}
\left({\partial_t}+\myvec\nabla_p H\cdot\myvec\nabla_r
-\myvec\nabla_r H\cdot\myvec\nabla_p\right)F_i(\myvec r, \myvec p,
\mu; t) = I_{\rm coll}[F_i,F_N,F_\pi,F_{\Delta},...]. \nonumber
\end{equation}
This equation describes the space-time evolution of the
generalized phase space density $F_i(\myvec r, \myvec p; t)$ of
particles of type $i$ with invariant mass $\mu$ under the
influence of the Hamiltonian $H$. The BUU equations are coupled
via the mean field in $H$ and via the collision term $I_{\rm
coll}$. The collision integral $I_{\rm coll}$ accounts for elastic
and inelastic collisions, decays and the formation of resonances,
including Pauli blocking. FSI therefore lead to absorption, charge
exchange, a redistribution of energy and to the production of new
particles.

\begin{figure}
        \centerline{
                \includegraphics[scale=\doubleplotscale]{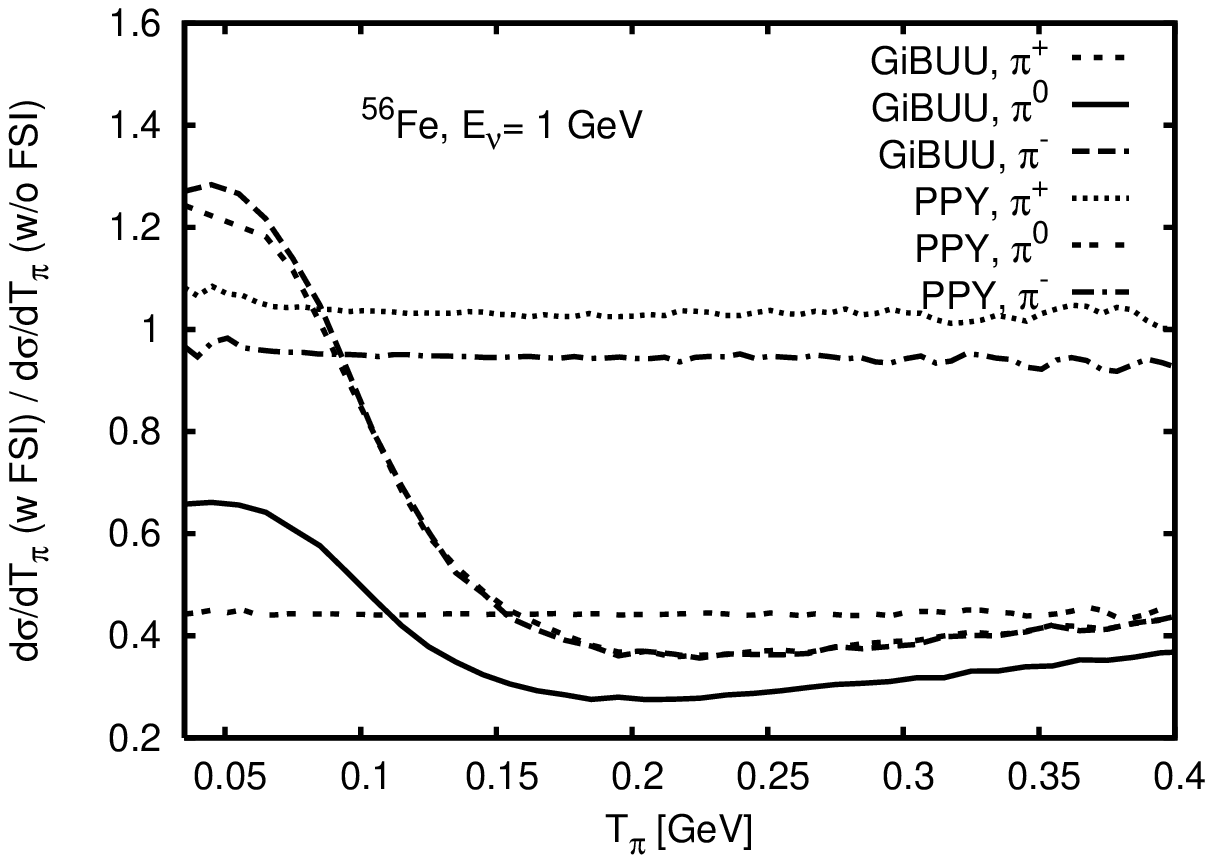}
                \hspace{1em}
                \includegraphics[scale=\doubleplotscale]{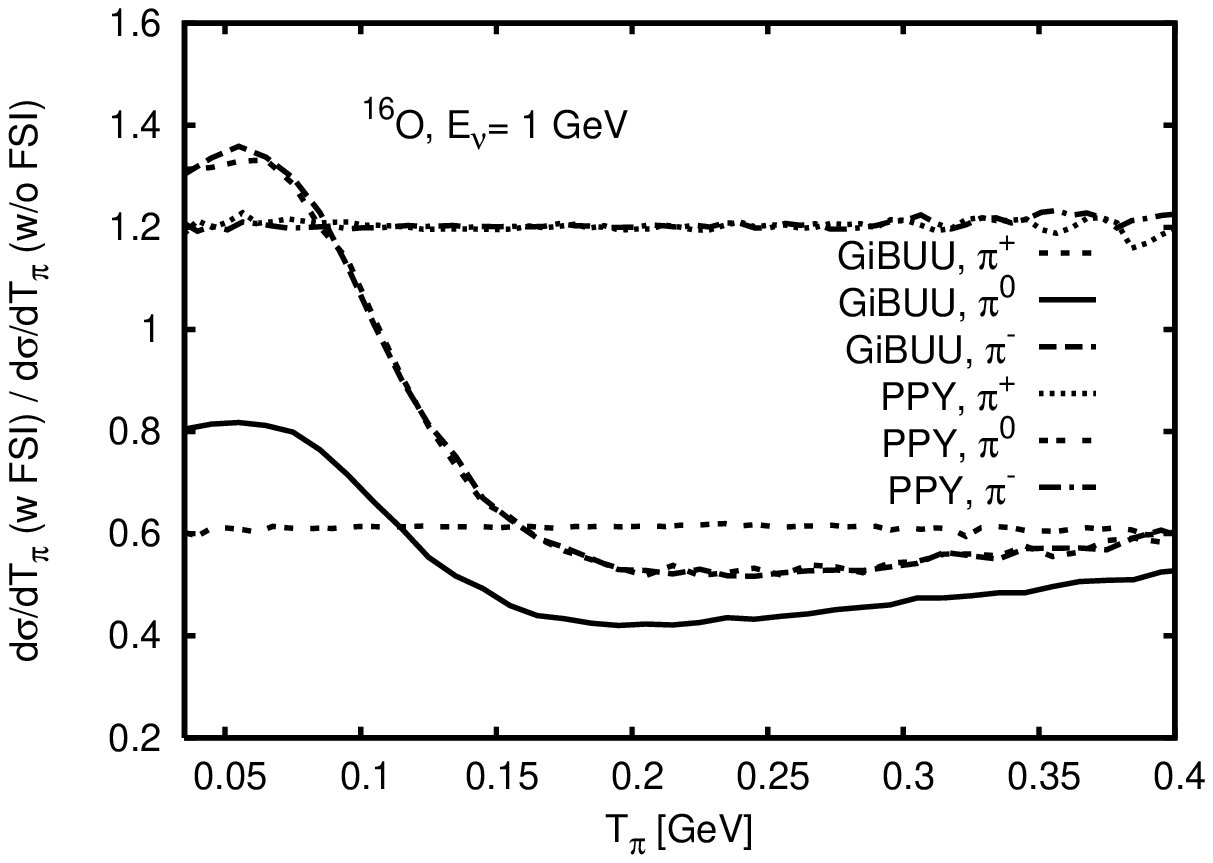}
        } \caption{Ratio of the NC differential cross section (the
        cross section with FSI divided by the one without FSI) for
        NC pion production on $^{56}$Fe (left) and
        $^{16}$O (right) versus the pion kinetic energy for
        $E_{\nu}= 1 \myunit{GeV}$. The initial QE scattering
        process has been "switched off". The curves labeled "PPY"
        denote the results of Paschos \refetal{Paschos:2000be,
        Paschos:2007pe}. \label{fig:pion_production}}
\end{figure}

The impact of FSI effects on NC induced pion production is clearly
visible in the ratios obtained by dividing the kinetic energy
spectra with FSI by the one without FSI (cf.~curves in
\reffig{fig:pion_production} labeled "GiBUU"; see
\refcite{Leitner:2006} for details). The absorption is bigger for
$^{56}$Fe (left panel) than for $^{16}$O (right panel), as
expected. For pions with kinetic energy above~$\approx 0.1$~GeV we
find large effects of FSI, with especially strong suppression
around $T_{\pi} \approx 0.13$~GeV. This is the region where pion
absorption and rescattering are most prominent due to the
excitation of the $\Delta$ resonance ($\pi N \to \Delta$ followed
by $\Delta N \to N N$, $\Delta N N \to N N N$, $\Delta N \to \pi N
N$ or $\Delta N \to \Delta N$). At lower pion energies we find a
peak because pions of higher energy in average loose energy via
$\pi N$ rescattering. Since side-feeding shifts strength always
from the dominant into the less dominant channel we find that FSI
lead to a strong reduction of the total yield in the
$\pi^0$~channel (solid line) while the reduction is much smaller
in the $\pi^+$ and $\pi^-$~channels (dashed lines); there the
ratio even exceeds 1 at low kinetic energies. A similar pattern
has been experimentally observed in pion photoproduction
(cf.~Fig.~16 in \refcite{Krusche:2004uw}). The particular
dependence of both ratios reflects well-known features of the $\pi
N \Delta$ dynamics in nuclei.

The curves in \reffig{fig:pion_production} labeled with "PPY" give
the corresponding results obtained from a calculation of Paschos
\refetal{Paschos:2000be,Paschos:2007pe}. In this article we
compare with the corrected results of
\refcite{Paschos:2007pe}\footnote{The original
work~\cite{Paschos:2000be} has an error in the elementary pion
production cross section.} where we divided their "ig" result
(with only Pauli-blocking) by their "f" result (with all nuclear
corrections) and rescaled $E_{\pi}$ to $T_{\pi}$. We now focus on
the comparison of both FSI models, the ANP
model~\cite{Adler:1974qu} used by Paschos~\emph{et al.}~and our
GiBUU model. We find that both FSI models are quantitatively and
even qualitatively very different ("GiBUU" vs. "PPY" curves). The
ratios of Paschos \emph{et al.}~for the charged pions are
considerably larger than ours for kinetic energies $> 0.1$~GeV. In
addition, they are practically flat as a function of the pion
energy, in contrast to our results. In our calculation the ratio
is much larger at low pion energies than at the higher ones
because pions rescatter with the nucleons (with or without charge
exchange) in the nuclear medium; in doing so they loose energy.
After the first collision, due to the energy redistribution, the
probability of a second collision changes. The ANP model, on the
contrary, assumes that the energy of the pion is constant during
its random walk through the nucleus. Also, the ANP model uses
vacuum cross sections to estimate the collision probability
ignoring in-medium modifications. This is especially important for
pions in the $\Delta$ region since this resonance is considerably
broadened in the medium.

\begin{figure}
        \centerline{
                \includegraphics[scale=\doubleplotscale]{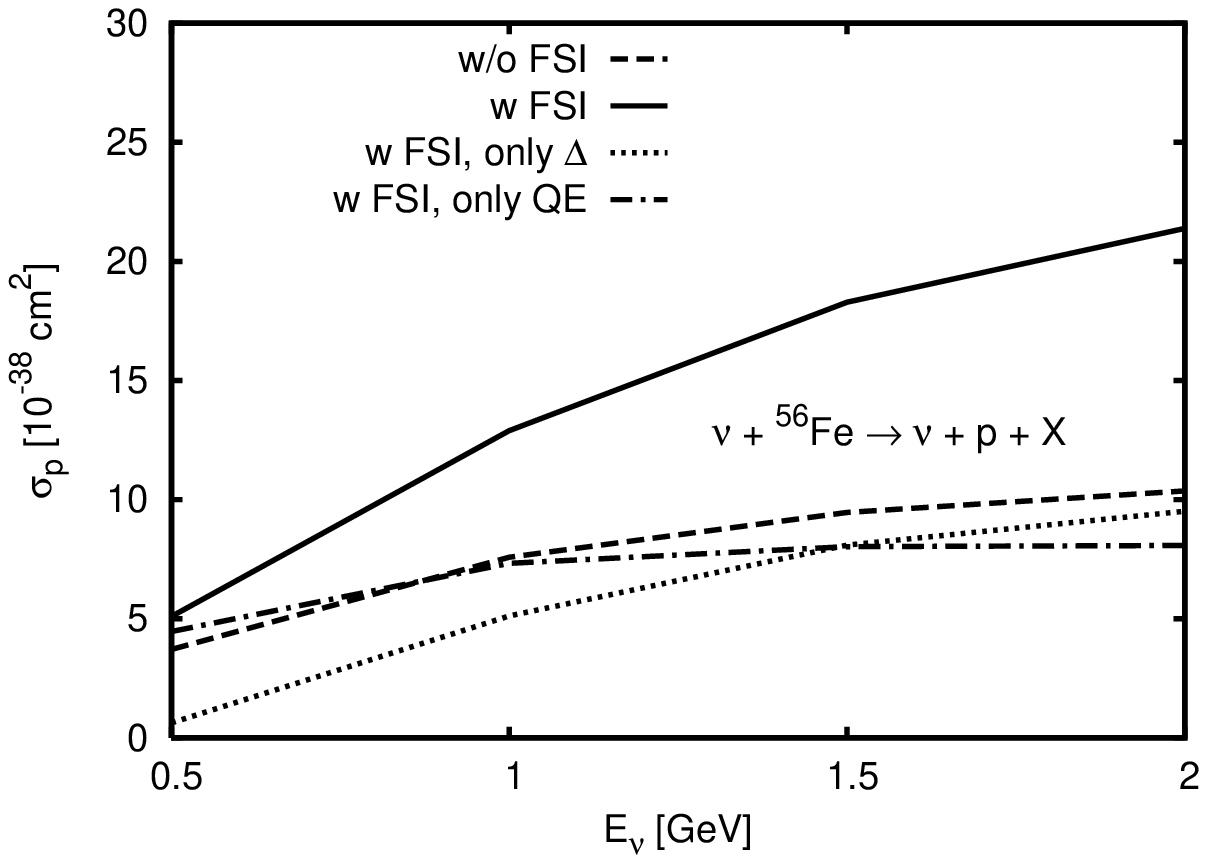}
                \hspace{1em}
                \includegraphics[scale=\doubleplotscale]{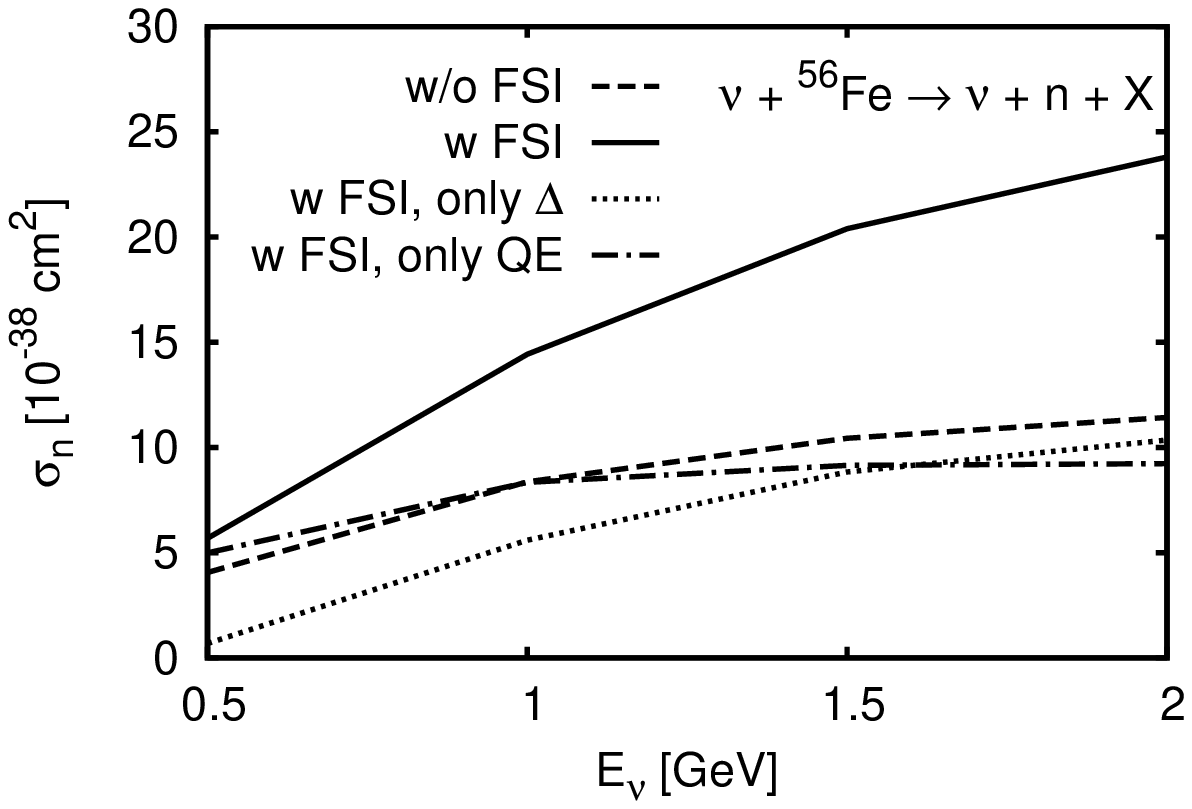}
        } \caption{Integrated cross section for NC induced proton
        (left) and neutron (right) knockout on $^{56}$Fe versus
        $E_{\nu}$. The dashed lines show the results without FSI;
        the results denoted by the solid lines include FSI. Also
        indicated is the contribution of QE (dash-dotted) and
        $\Delta$ excitation (dotted) to the total yield. Multi-nucleon
        knockout is taken into account.
        \label{fig:nucleon_knockout}}
\end{figure}

A correct understanding of the in-medium $\pi N \Delta$ dynamics
is crucial for the interpretation of experiments to ensure proper
identification of QE events. This is visualized in
\reffig{fig:nucleon_knockout} where we show the NC induced cross
sections for proton and neutron knockout on $^{56}$Fe. The solid
lines, showing the results with FSI included, lie in both cases
clearly above the ones without FSI (dashed lines); this
enhancement is caused by secondary interactions. Furthermore, it
is indicated whether the knockout was induced by initial QE
scattering (dash-dotted) or $\Delta$ excitation (dotted) through
e.g.~$\Delta N \to N N$, $\Delta \to \pi N$, $\Delta N N \to N N
N$. Both contribute to the cross section above $E_{\nu} \approx
1.2$~GeV with almost equal amounts. If an initially produced pion
gets absorbed, then this event looks "QE-like" which can lead to
systematic errors in the data analysis. This is particularly
critical for the NC case, where the outgoing neutrino remains
undetected. Only for neutrino energies up to $\approx 0.5$~GeV one
can neglect the resonance contributions to nucleon knockout.

%\section{Summary and Outlook}

We conclude that with the present knowledge of $\pi N \Delta$
dynamics in nuclei, based on extensive studies of pion and
photo-nuclear reactions, a realistic, quantitative description of
neutrino induced pion production in nuclei is possible. In-medium
effects in $\nu A$ scattering, and in particular FSI, are
important for the interpretation of LBL oscillation experiments.
The influence of final state interactions, therefore, has to be
treated with the same degree of sophistication as the primary
production process and the nuclear spectral function information.

%%%%%%%%%%%%%%%%%%%%%%%%%%%%%%%%%%%%%%%%%%%%%%%%
%% BACKMATTER
%%%%%%%%%%%%%%%%%%%%%%%%%%%%%%%%%%%%%%%%%%%%%%%%

%\begin{theacknowledgments}
We thank all members of the GiBUU project for cooperation. This
work has been supported by the Deutsche Forschungsgemeinschaft and
by BMBF.
%\end{theacknowledgments}

\end{document}